\documentclass[aps,
 prd,
 reprint,
 raggedbottom,
 superscriptaddress,
 nofootinbib,
 nobibnotes,
 amsmath,amssymb,
 nolongbibliography,
 noeprint
]{revtex4-2}

\usepackage{graphicx}
\usepackage{dcolumn}
\usepackage{bm}
\usepackage[pdftex,colorlinks=true,linkcolor=blue,citecolor=blue,urlcolor=blue]{hyperref}
\makeatletter
\newcommand{\fixedfigures}{%
  \let\PRD@savedfigure\figure
  \let\PRD@savedendfigure\endfigure
  \renewcommand\figure[1][]{%
    \par\noindent\begin{minipage}{\linewidth}%
    \def\@captype{figure}%
    \centering
  }%
  \renewcommand\endfigure{\end{minipage}\par}%
}
\newcommand{\restorefigures}{%
  \let\figure\PRD@savedfigure
  \let\endfigure\PRD@savedendfigure
}
\makeatother

\begin{document}

\title{QPEs from Warped Disk Collisions with EMRIs: Brightness Recurrence Correlation and Gravitational-Wave Follow-up}


\author{Bo-An Chen}
\affiliation{Department of Astronomy, School of Physics and Technology, Wuhan University, Wuhan 430072, China}

\author{Bei You}
\email[Contact author: ]{youbei@whu.edu.cn}
\affiliation{Department of Astronomy, School of Physics and Technology, Wuhan University, Wuhan 430072, China}

\author{Giovanni Miniutti}
\affiliation{Centro de Astrobiología (CAB), CSIC-INTA, Camino Bajo del Castillo s/n, 28692 Villanueva de la Cañada, Madrid, Spain}

\author{Ning Jiang}
\affiliation{Department of Astronomy, University of Science and Technology of China, Hefei 230026, China}
\affiliation{School of Astronomy and Space Sciences, University of Science and Technology of China, Hefei 230026, China}

\author{Zhen Pan}
\affiliation{Tsung-Dao Lee Institute, Shanghai Jiao-Tong University, 1 Lisuo Road, Shanghai 201210, China}
\affiliation{School of Physics and Astronomy, Shanghai Jiao-Tong University, 800 Dongchuan Road, Shanghai 200240, China}

\author{Tao Yang}
\affiliation{Department of Astronomy, School of Physics and Technology, Wuhan University, Wuhan 430072, China}

\author{Xi-Long Fan}
\affiliation{Department of Astronomy, School of Physics and Technology, Wuhan University, Wuhan 430072, China}

\author{Kai Liao}
\affiliation{Department of Astronomy, School of Physics and Technology, Wuhan University, Wuhan 430072, China}

\author{Xu-Heng Ding}
\affiliation{Department of Astronomy, School of Physics and Technology, Wuhan University, Wuhan 430072, China}

\author{Zong-Hong Zhu}\email[Contact author: ]{zhuzh@whu.edu.cn}
\affiliation{Department of Astronomy, School of Physics and Technology, Wuhan University, Wuhan 430072, China}
\affiliation{Department of Astronomy, Beijing Normal University, Beijing 100875, China}

\author{Shuai-Kang Yang}
\affiliation{Department of Astronomy, School of Physics and Technology, Wuhan University, Wuhan 430072, China}

\author{Sai-En Xu}
\affiliation{Department of Astronomy, School of Physics and Technology, Wuhan University, Wuhan 430072, China}

\author{Han He}
\affiliation{Department of Astronomy, School of Physics and Technology, Wuhan University, Wuhan 430072, China}

\author{Xiao Fan}
\affiliation{Department of Astronomy, School of Physics and Technology, Wuhan University, Wuhan 430072, China}


\date{\today}

\begin{abstract}
Quasi-Periodic Eruptions (QPEs) display correlated long/short and strong/weak patterns that remain unexplained by existing flat-disk collision models. 
We propose that these features arise from an extreme-mass-ratio inspiral (EMRI) colliding with a warped accretion disk, likely formed after a tidal disruption event. The warp modulates both recurrence time and burst energy, encoding the disk geometry\textemdash and thus the spin of the central supermassive black hole (SMBH)\textemdash into the X-ray light curve. We introduce the Brightness-Recurrence Diagram (BRD) to visualize this correlation, where QPE bursts trace an elliptical trajectory driven by the EMRI's apsidal precession; the tilt of this ellipse encodes whether the EMRI is prograde or retrograde relative to the SMBH spin. Applying this model to the prototypical QPE source GSN~069 successfully reproduces the observed patterns. The data are consistent with either a prograde stellar secondary or a retrograde stellar-mass black hole. In the stellar-mass black hole scenario, ongoing orbital decay could render the EMRI detectable by LISA within a few decades, facilitating gravitational-wave follow-up and independent multimessenger constraints on the system.
\end{abstract}


\maketitle

\section{Introduction}
\label{sec:intro}
Quasi-periodic eruptions (QPEs) are luminous X-ray bursts from galactic nuclei, recurring on timescales of hours to more than 10 days.
The first QPE was discovered in 2019 from the nucleus of GSN~069.
Since then, QPEs have emerged as an important topic in high-energy astronomy, with more than 10 QPE sources observed to date \citep{2019Natur.573..381M, 2013ApJ...768..167S, 2020A&A...636L...2G, 2021Natur.592..704A, 2024A&A...684A..64A, 2025ApJ...989...13A, 2021ApJ...921L..40C, 2024Natur.634..804N, 2025NatAs...9..895H, 2025ApJ...983L..39C, 2025MNRAS.540...30B,2025A&A...703A.263H,2026A&A...706L..15B}.

The mechanism behind QPEs remains uncertain. Proposed models include collisions between an extreme-mass-ratio inspiral (EMRI) and the supermassive black hole's (SMBH's) accretion disk \citep{2010MNRAS.402.1614D,2021ApJ...921L..32X,2023ApJ...957...34L, 2023A&A...675A.100F,2023MNRAS.526...69T,2024PhRvD.109j3031Z,2024PhRvD.110h3019Z,2025ApJ...978...91Y,2025ApJ...985..242Z,2026ApJ..1000L..57G,2026ApJ...1003..148A}, accretion disk instabilities \citep{2023A&A...672A..19S,2022ApJ...928L..18P,2023ApJ...952...32P,2021ApJ...909...82R,2025ApJ...989..196P}, mass transfer onto a SMBH from an orbiting body \citep{2020MNRAS.493L.120K,2022MNRAS.515.4344K,2023MNRAS.520L..63K,2022A&A...661A..55Z,2023ApJ...947...32C,2024MNRAS.529.1440W,2022ApJ...933..225W,2025MNRAS.543.3503Y}, gravitational lensing of a SMBH binary \citep{2021MNRAS.504.5512I}. 

Among these models, the EMRI-disk collision model is particularly adept at explaining 
the regular alternation between long and short recurrence times ($T_{\rm long}$/$T_{\rm short}$)
observed in sources such as GSN~069. However, the situation becomes more complex when the strong/weak and long/short properties are examined together. Observations show that these two characteristics are systematically correlated: in GSN~069, strong bursts are generally preceded by short intervals (Fig.~4 in \citet{2023A&A...670A..93M}), whereas the opposite trend is seen in RX~J1301.9+2747 (Fig.~2 in \citet{2024A&A...692A..15G}). 

This correlation is readily apparent by eye and strangely long-lived.
In GSN~069, multiple QPE light-curve segments spanning more than four years (2019--2023) all maintain the correlation.
The eruptions even became vanished during a possible second TDE, yet the QPEs that later reappeared seem to preserve the pattern of long intervals following strong bursts and short intervals following weak bursts \citep{2023A&A...670A..93M,2025A&A...693A.179M}.
In RX~J1301.9+2747, a similar analysis of light-curve segments from 2000, 2019, 2020, and 2022 likewise finds a long-lived correlation, but with the opposite trend \citep{2024A&A...692A..15G}.

In both sources, the correlation has persisted over a timescale clearly longer than the apsidal precession period $T_{\rm aps}$ inferred in collision models: Bayesian timing analyses give $T_{\rm aps}\sim76$~d for GSN~069 \citep{2025MNRAS.543.1816Z}, while $T_{\rm aps}$ is even shorter for RX~J1301.9+2747 \citep{2024A&A...692A..15G,2023A&A...675A.100F,2024PhRvD.110h3019Z}.

In contrast, in the collision model, apsidal precession implies that stronger bursts are equally likely to be preceded by either long or short intervals\textemdash a prediction inconsistent with observations. This discrepancy suggests that an additional physical ingredient is missing from the collision model.

Another piece of evidence supporting the EMRI-disk collision model is that approximately half of known QPEs are observed years after tidal disruption events (TDEs), implying the presence of a TDE disk in these sources \citep{2018ApJ...857L..16S,2021ApJ...920L..25S,2023A&A...675A.152Q,2025ApJ...983L..18J}.
The collision models generally assume a flat (and sometimes rigidly precessing) disk, a geometry commonly adopted in studies of the early evolution of inclined TDE disks \citep{2016MNRAS.455.1946F,2016MNRAS.463.2242X,2018MNRAS.481.3470I,2025SciA...11y9068W}.
However, at the late stage of TDE evolutions, the accretion disk is expected to be thin and inclined, implying that a Bardeen-Petterson warp structure should be taken into account \citep{2025ApJ...985..146G,2025ApJ...992..114G,2012PhRvL.108f1302S,2023ApJ...957L...9T}. 

\begin{figure}[tbp]
\includegraphics[width=0.45\textwidth]{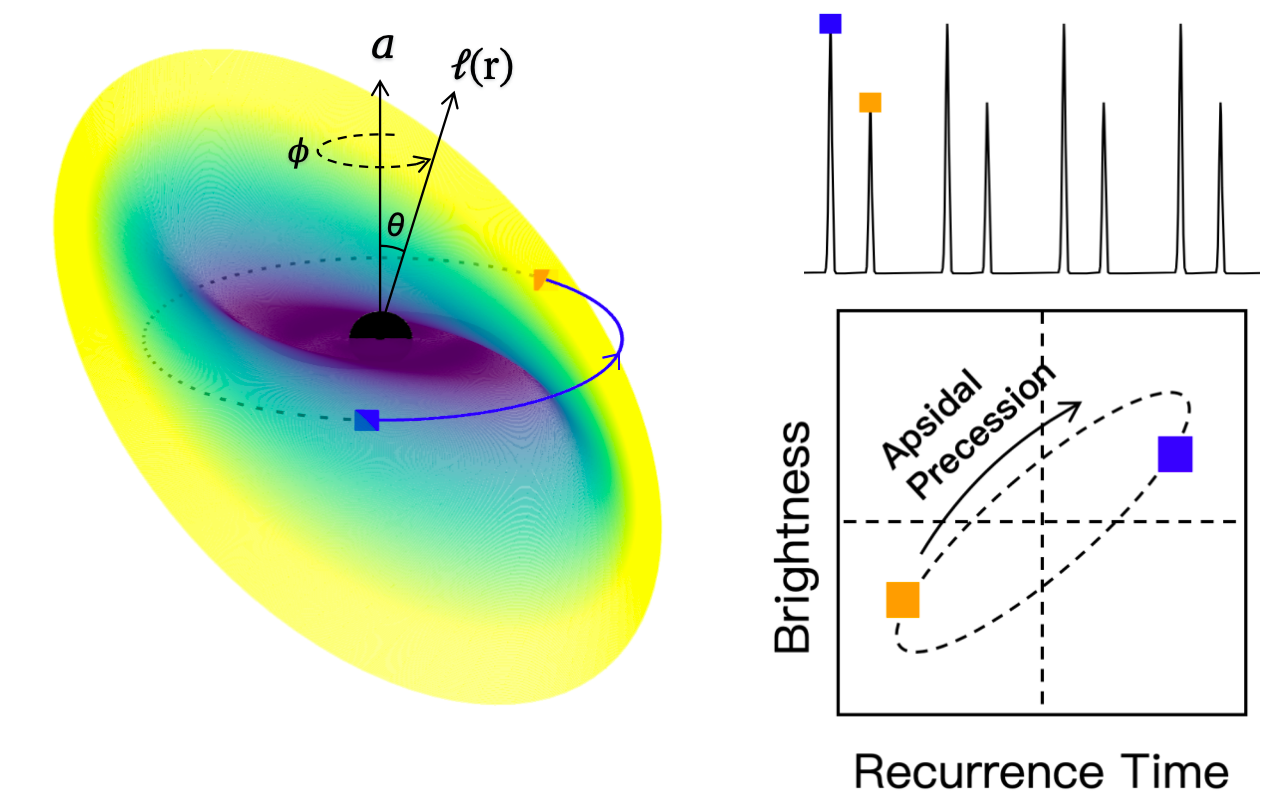}
\caption{The correlated long/short and strong/weak patterns in QPEs arise from the collision between a stellar-mass black hole on an eccentric orbit and a warped accretion disk. The modulation is driven by the varying impact distance, which imprints the warp geometry (hence the SMBH's spin) onto the Brightness-Recurrence Diagram.}
\label{fig:geometry}
\end{figure}

In this paper, we demonstrate that a Bardeen-Peterson warp structure could produce correlated long/short and strong/weak QPE patterns. 
In this scenario, the QPE light curves carry imprints of the warp geometry linked to the central black hole's spin, revealing at least whether the spin is prograde or retrograde with respect to the EMRI orbit. 
Motivated by recent observational studies \citep{2023A&A...670A..93M,2023A&A...674L...1M,2024A&A...692A..15G}, we construct the Brightness-Recurrence Diagram (BRD) to analyze how the observed X-ray QPEs encode information about an EMRI system. See Fig.~\ref{fig:geometry} for a detailed schematic illustration.
Throughout this paper, we take the prototypical TDE-associated QPE source GSN~069 as our primary example to develop and test this scenario \citep{2019Natur.573..381M,2023A&A...670A..93M,2023A&A...674L...1M,2025A&A...693A.179M}.

Observations further indicate that the EMRI orbital period in several QPEs is decaying \citep{2023A&A...670A..93M,2025ApJ...985..242Z,2024A&A...690A..80A, 2025arXiv250807961L}. If the secondary object is a stellar-mass black hole, these EMRIs are expected to enter the sensitivity band of space-based gravitational-wave (GW) detectors such as LISA, TianQin, and Taiji \citep{2024arXiv240207571C,2016CQGra..33c5010L,2020IJMPA..3550075R} within a few decades, enabling GW follow‑up. A multimessenger detection linking a GW EMRI with its electromagnetic QPE counterpart is therefore of great interest.
Such multimessenger EMRI candidates could be confirmed through at least three independent lines of verification: (1) the association of EM with GW in terms of the spatial localization and the consistency between the host galaxy redshift and the GW luminosity distance; (2) constraints on the central black hole mass \citep{2025MNRAS.543.1816Z}; (3) the spin of the central black hole (as addressed in our work). Once such a target is identified, it would offer rich scientific potential, enabling, for example, precise multimessenger parameter estimation, constraining cosmology
\citep{2021MNRAS.508.4512L,2026ApJ...997..134Z}, and probing fundamental physics with improved parameter constraints \citep{2017PhRvD..95j3012B,2022NatAs...6..464M,2024arXiv240108085C}.
Such systems may also help constrain the milli-Hertz gravitational-wave background \citep{2022ApJ...930..122C}.



\section{Warped Disk Structure}
\label{sec:warped}
The evolution of a warped disk can be described by \citep{2000ApJ...538..326L,1995ApJ...438..841P,2025ApJ...978..103P} 
\begin{equation}\label{eq:warp}
    \Sigma r^{2} \Omega \frac{\partial W}{\partial t}=\frac{1}{r} \frac{\partial G}{\partial r}+T,
\end{equation}
where $W(r,t)=\hat{\boldsymbol{l}}(r,t)\cdot(\hat{\boldsymbol{x}}+\mathrm{i}\hat{\boldsymbol{y}})$ is the complex warp amplitude of the disk; ${\boldsymbol{l}}$ is the specific angular momentum; $\Sigma$ is the surface density and we set $\Sigma \propto r^{-3/5}$ \citep{1973A&A....24..337S}; $\Omega$ is the orbital frequency. $G$ and $T$ are the complex internal and external torques, respectively.

The external torque $T=\mathrm{i} \Sigma r^{2} \Omega \Omega_{\mathrm{LT}} W$ arises from the Lense-Thirring (LT) effect.
The LT frequency is
\begin{equation}\label{eq:LT_frequency}
    \Omega_{\rm LT}=\frac{c^3}{GM}\frac{1}{r^{3/2}+a}\left( \frac{2a}{r^{3/2}}+\frac{3a^2}{2r^2}\right),
\end{equation}
where $a$ is the spin parameter of the central black hole.

The form of the internal torque depends on the disk thickness. For a sufficiently thin disk ($H/R < \alpha$), the inner region is expected to be aligned to the SMBH's equatorial plane (the so-called Bardeen-Petterson effect \citep{1975ApJ...195L..65B}).
Assuming a steady disk where the radial velocity $v_r = {3\alpha H^2 \Omega}/{2r}$, the internal torque is given by \citep{1999MNRAS.304..557O}
\begin{equation}\label{eq:internal_torque}
    G=\Sigma H^{2} r^{3} \Omega^{2}\left(Q_{\mathrm{v}} \frac{\partial W}{\partial r}+\mathrm{i} Q_{\mathrm{p}} \frac{\partial W}{\partial r}\right),
\end{equation}
where $Q_{\rm v}$ and $Q_{\rm p}$ are the viscous and pressure coefficients \citep{1999MNRAS.304..557O,2019MNRAS.487.4965Z}; $H$ is the disk's half-thickness.

\begin{figure}[tbp]
\includegraphics{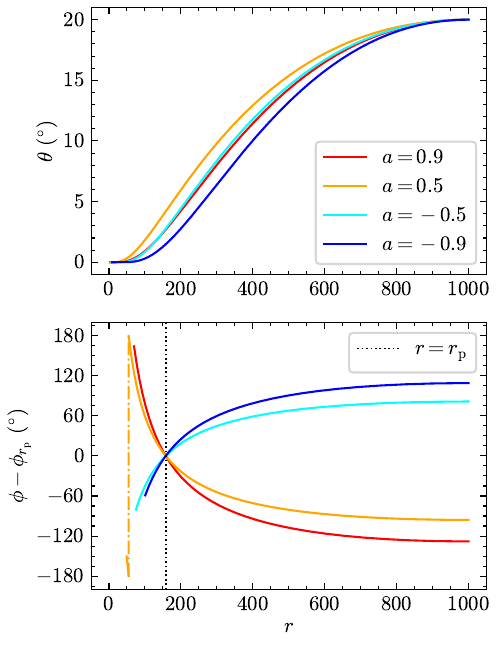}
\caption{Structure of a warped TDE disk. The upper panel shows the radial profile of the disk inclination angle $\theta$. The lower panel displays the radial variation of the twist angle $\phi$. Parameters are set to $M_\bullet=10^{5.99}M_\odot$, $r_{\rm out} = 1000 M_\bullet$, $\theta_{\rm out}=20^\circ$, $H/R=0.0055$, $\alpha = 0.1$, $r_{\rm p}=157.7M_\bullet$ denotes the characteristic pericenter distance for the EMRI in GSN~069. 
}
\label{fig:warpeddisk}
\end{figure}

\label{sec:disk-derivation}
\label{SM:Derivation}
We solve the equations by separation of variables and seek the lowest-order eigenmode solution \citep{2019MNRAS.487.4965Z}, writing
\begin{equation}
  \begin{aligned}
    W(r,t) &= \tilde{W}(r)\,\mathrm{e}^{\int\lambda\,\mathrm{d}t} \\
    G(r,t) &= \tilde{G}(r)\,\mathrm{e}^{\int\lambda\,\mathrm{d}t},
  \end{aligned}
\end{equation}
where $\lambda = \gamma+\mathrm{i}\omega$ is the complex eigenfrequency, with $\omega$ and $\gamma$ representing the precession and damping rate of the warped disk.
Here $\lambda$ is a constant over time for a steady disk, implying the background disk properties change over time-scales longer than $|\lambda|^{-1}$.
The boundary conditions are
$
    \tilde{G}\left(r_{\text {in}}\right)=\tilde{G}\left(r_{\text {out}}\right)=0,
$
where $r_{\text{in}}$ is the innermost stable circular orbit \citep{1972ApJ...178..347B}.
For GSN~069, QPEs appear roughly eight years after the TDE, well beyond the stream-fed phase; a finite-mass disk with a free outer boundary $\tilde{G}(r_{\rm out})=0$ is therefore appropriate.
We employ a shooting method to determine the lowest-order eigenmode of the warped disk; higher-order modes, with larger damping rates, are negligible.

Related studies of TDE-disk warping have focused on the early evolution under combined stream and Lense--Thirring torques \citep{2016MNRAS.463.2242X,2018MNRAS.481.3470I}; here we instead adopt a late-stage thin-disk description and solve for the quasi-steady Bardeen--Petterson eigenmode relevant to subsequent EMRI--disk collisions.
We do not claim a detailed model of the transition from a stream-fed, continuously evolving disk to a finite-mass one; so long as the late disk is not fully aligned, the eigenmode decomposition ensures that the lowest-order mode ultimately dominates.
We also notice that Teboul \& Metzger \citep{2023ApJ...957L...9T} found that for a spreading disk the alignment can start at the spreading radius, giving rise to a warped disk; exploring that scenario may be left for future work.

The typical warp structure is quantified by the complex warp amplitude $\tilde{W}(r) = e^{\mathrm{i}\phi} \sin{\theta}$, where $\theta=\arcsin|\tilde{W}|$, $\phi=\arg\tilde{W}$ are the inclination angle and the twist angle, respectively.
A positive black hole spin results in a phase lead of the twist angle in the inner disk region, while a negative spin produces a lag because the sign reversal of the LT torque. The disk does align with time: for our fiducial parameters ($r_{\rm out}=1000\,R_g$, $\alpha=0.1$, $H/R=0.0055$), the lowest-order eigenmode gives $t_{\rm prec}=1/\omega\sim20~\mathrm{yr}$ and $t_{\rm align}=1/\gamma\sim33~\mathrm{yr}$, much longer than hundreds of QPE cycles yet much shorter than the viscous time at the outer edge, $t_{\rm visc}\sim r^{2}/(\alpha\Omega H^{2})\sim1.6\times10^{3}~\mathrm{yr}$. A quasi-steady warp (and a steady $v_r$ in the internal torque) is therefore a reasonable late-time assumption, and we neglect precession and alignment of the disk over the QPE observing window. We therefore also expect the warp geometry, and the QPE pattern it produces, to remain roughly unchanged for up to about a decade, provided the accretion disk remains in the thin-disk regime.

These disk assumptions are also consistent with observational constraints on the TDE disk in GSN~069 \citep{2025ApJ...985..146G,2025ApJ...992..114G}.

Typical late-stage TDE warp structure (Fig.~\ref{fig:warpeddisk}) shows smaller inner-disk inclination and leading/lagging twist phase. This warp structure modulates the EMRI-disk intersection points, naturally producing the observed QPE patterns (see below).

\section{QPE Patterns from Warped Disk Collisions}
\label{sec:patterns}
Given a slightly eccentric orbit,  the long/short pattern is modulated by the warp geometry through radial variations in the twist angle. 
A possible estimate is that, for an EMRI with a semi-major axis of \(A\sim160 M_\bullet\) and an eccentricity of \(e\sim0.04\) (commonly inferred for GSN~069 \citep{2025ApJ...983L..18J}), as it crosses a warped disk characterized by Fig.~\ref{fig:warpeddisk}, the twist angle changes by \(\delta \phi = \phi(r_{\rm p})-\phi(r_{\rm a})\approx\) \(3.4^{\circ}\), where $r_{\rm p}$ and $r_{\rm a}$ are the pericenter and apocenter distances, respectively. This contributes to a long/short recurrent pattern with \({\delta t_{\rm rec}}/{t_{\rm rec}}\approx 2{\delta \phi}/{180^\circ}\approx\) \(4\%\). Since both the disk warp and the orbital eccentricity contribute to the observed long/short pattern, the eccentricity required to match the data would be lower than the nominal value of \(e\sim0.04\) inferred under the assumption of a flat disk. For the warped disk, we infer a lower eccentricity \(e\sim0.015\).

The strong/weak modulation reflects the variation in the radiative energy. 
However, the radiative transfer process behind the QPEs is far more uncertain, in this work we simplify the collision energy as follows. 
Considering a black hole as the secondary object \citep{2023A&A...675A.100F}, the collision energy is estimated as \citep{2024PhRvD.109j3031Z}
\begin{equation}\label{eq:collision_energy_bh}
E_{\rm sBH} =\frac{4 \pi}{v_{\mathrm{rel}}^{2}} G^{2} m_{\rm sBH}^{2} \Sigma \ln\Lambda \propto v_{\rm rel}^{-2}\ ,
\end{equation}
where $v_{\mathrm{rel}}$ denotes the relative speed between the secondary and the disk gas, and $\ln \Lambda$ is the Coulomb logarithm. 
Consider collisions between an equatorial EMRI and the warped disk, the angle between the EMRI orbital plane and the local disk surface is simply $\theta$ 
and the relative velocity scales as $v_{\mathrm{rel}}\propto\theta$. The typical variation in $\theta$, quantified as $1-\theta(r_{\rm p})/\theta(r_{\rm a}) \sim$ 8\% (Fig.~\ref{fig:warpeddisk}), leads to a roughly 16\% variation in collision energy, thereby producing the strong/weak modulation. The derived characteristic long/short and strong/weak modulations are consistent with the observed behavior of QPEs like GSN~069.

It is worth noting that the viability of the sBH–disk collision model depends on specific parameter choices. Early studies have shown that for such a scenario to work, the secondary's orbit must be prograde with respect to the gas flow (with a relative inclination $\lesssim 10^\circ$), and its mass must be relatively high, typically $\sim 100\,M_\odot$ \citep{2023A&A...675A.100F,2025arXiv250421456M,2026arXiv260302302A}. 
However, recent simulations indicate that these constraints may be relaxed, as the effective radius of a sBH could have been underestimated \citep{2026arXiv260300226L}. Within such a framework, the scenario remains viable and is explored in this paper.

On the other hand, if the secondary object is a star, the collision energy is estimated as
\begin{equation}\label{eq:collision_energy}
E_{\star} = \pi v_{\mathrm{rel}}^2R_{\star}^2\Sigma \propto v_{\rm rel}^2\ .
\end{equation}
Radiative processes associated with star–disk collisions have been studied extensively \citep{2023ApJ...945...86L,2025ApJ...991..147L,2025arXiv250421456M,2026arXiv260202656J,2025ApJ...993..186H,2025ApJ...983...40V}. In many cases, impacts between the debris stripped off the star and the disk, rather than the star–disk impact itself, have been invoked as the origin of QPEs. Eq.~\eqref{eq:collision_energy} does not account for such complex mechanisms. In this work, however, our primary focus is on how the warped disk modulates the burst energies. Accordingly, Eq.~\eqref{eq:collision_energy} is intended only to capture the scaling \(E \propto v_{\rm rel}^2\).

In addition to the long/short and strong/weak patterns, the correlation between them have been observed in several QPE sources \cite{2023A&A...670A..93M,2024A&A...692A..15G} and can be naturally explained in this warped disk model.
Consider, for instance, a stellar-mass black hole on a prograde orbit relative to the SMBH spin: when the collision occurs closer to the central SMBH, the disk inclination is smaller (as the inclination at the inner radius is lower than that at the outer radius; see the upper panel in Fig.~\ref{fig:warpeddisk}). This leads to a lower relative velocity and hence a stronger burst (Eq.~\ref{eq:collision_energy_bh}). Meanwhile, the inner disk region leads in twist phase (as shown by the red curve in the lower panel of Fig.~\ref{fig:warpeddisk}), which postpones the collision timing. Therefore, a stronger burst is generally preceded by a longer recurrence interval. This follows naturally from the warp geometry. Other scenarios, including flat or rigidly precessing disks, do not predict a systematic correlation between the burst intensity and the recurrence time. 

Moreover, in the case of a retrograde EMRI orbit with the SMBH spin ($a < 0$), the phase relation of the twist angle is reversed: the inner disk region lags in phase (as represented by the blue curve in the lower panel of Fig.~\ref{fig:warpeddisk}). Therefore, a collision at a smaller radius tends to occur earlier, resulting in a stronger burst being preceded by a shorter recurrence interval. Thus, the sign of the correlation (positive or negative) between brightness and recurrence time encodes the prograde or retrograde nature of the EMRI system (here and in the following, “prograde/retrograde” refers to the relative orientation between the EMRI orbit and the SMBH spin).

\section{Brightness-Recurrence Diagram}
\label{sec:brd}
 
\begin{figure*}[tbp]
    \centering
    \begin{minipage}[b]{0.32\linewidth}
        \centering
        \includegraphics[width=\linewidth]{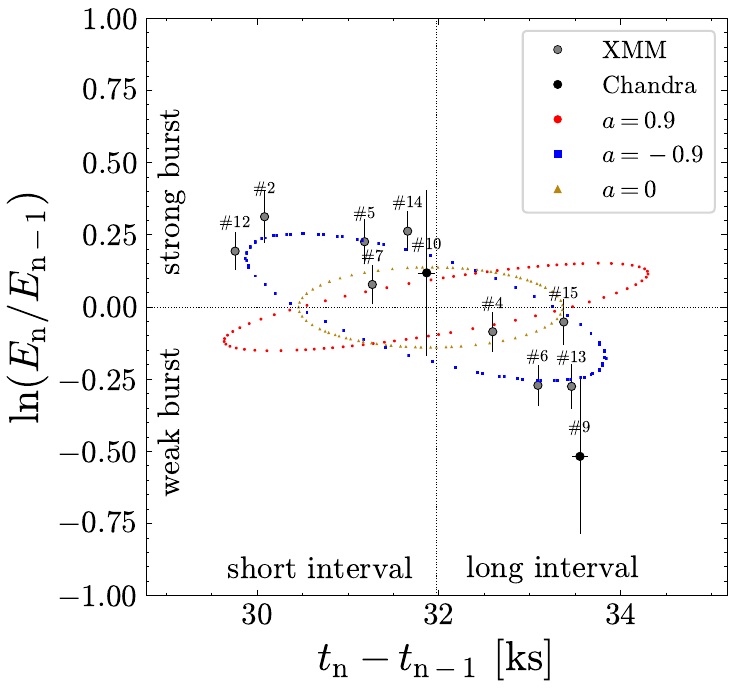}
        \text{(a) BRD for sBH secondary} 
    \end{minipage}\label{fig:A-T_warped_star}
    \hfill
    \begin{minipage}[b]{0.32\linewidth}
        \centering
        \includegraphics[width=\linewidth]{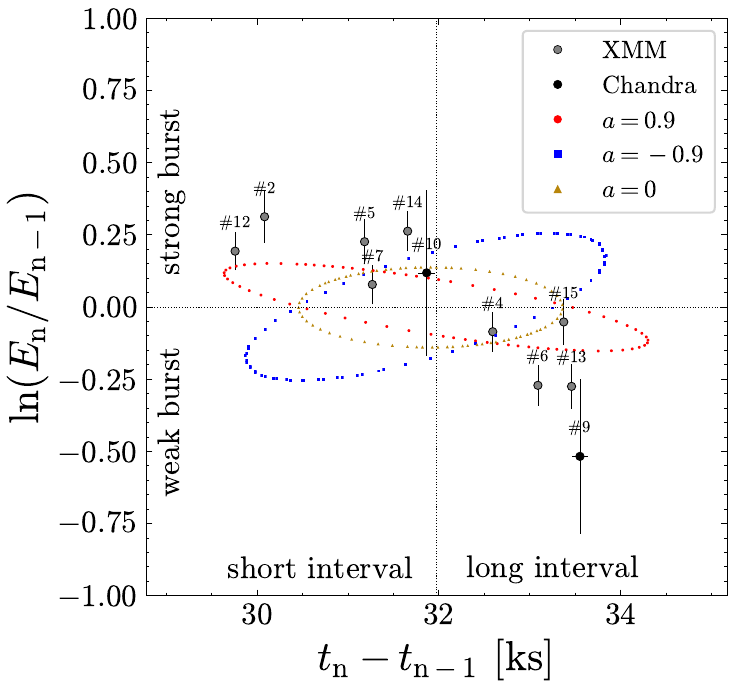}
        \text{(b) BRD for star secondary} 
    \end{minipage}\label{fig:A-T_warped}
    \hfill
    \begin{minipage}[b]{0.32\linewidth}
        \centering
        \includegraphics[width=\linewidth]{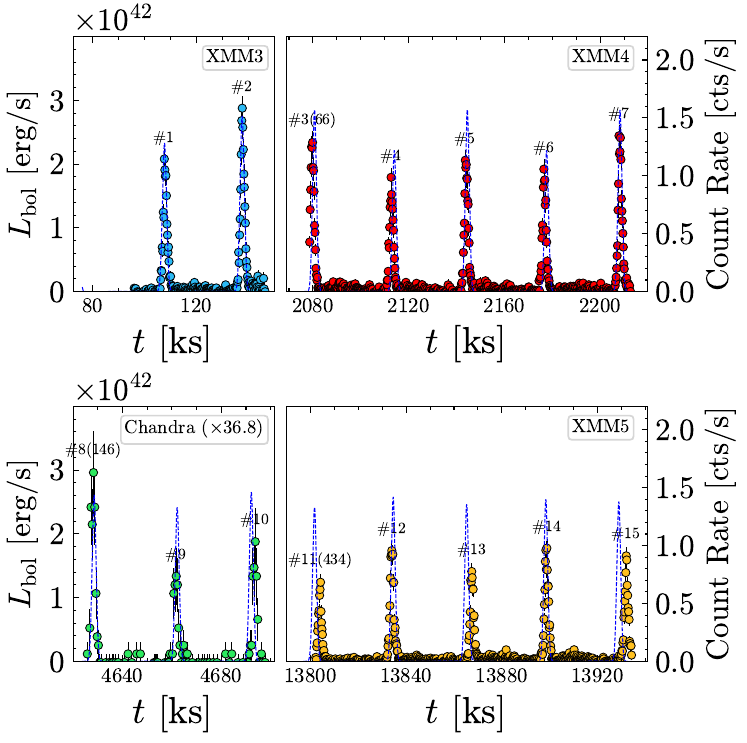}
        \text{(c) light curve for sBH secondary, $a<0$} 
    \end{minipage}\label{fig:Long-term}
    \caption{
    Brightness-Recurrence Diagram (BRD) and long-term light curve for GSN~069.
    (a) BRD with model predictions for positive (red dots), negative (blue squares), and zero (yellow triangles) spin scenarios, assuming a black hole secondary. Error bars show XMM (gray) and Chandra (black) data; Chandra measurements are highly uncertain due to low photon counts (\(\lesssim\) few tens per peak). The observed trend agrees with the negative spin case. The collision energy is inferred as \(E = \sqrt{2\pi}\,L_{\rm peak}\,\sigma\), where \(L_{\rm peak}\) is the bolometric peak luminosity from spectral fitting in the 0.2–1 keV band, and \(\sigma\) the Gaussian width, assuming each burst follows a Gaussian profile. The recurrence time is estimated as the time interval between successive peaks. The data points in this plane were plotted in Fig.~A1 of \citet{2023A&A...674L...1M}.
    (b) Same as (a), but for a star secondary. The observed trend agrees with the positive spin case.
    (c) Simulated (blue lines; $a=-0.9$, sBH secondary) and observed (colored error bars) QPE light-curve profiles \citep{2023A&A...670A..93M}. Peak labels indicate the observed order; parentheses denote the absolute order predicted by the model. The secular drop in absolute amplitude is not fitted (see text). A constant quiescent level has been subtracted.
    Parameters: \(M_\bullet = 10^{5.99}\,M_\odot\) \citep{2022A&A...659L...2W}, \(m_{\rm sBH} = 10\,M_\odot\), \(A = 160.1 M_\bullet\), \(e = 0.015\), \(r_{\mathrm{out}} = 1000 M_\bullet\) \citep{2025ApJ...985..146G,2025ApJ...992..114G}, \(\theta_{\rm out}=20^\circ\), \(H/R = 0.0055\), \(\alpha = 0.1\), \(\ln\Lambda = 10\), \(\Sigma = 10^5\,\mathrm{g/cm^2}\) at \(A\). 
    }
    \label{fig:fit_for_069}
\end{figure*}

We plot the BRD to further investigate the correlation.  
As shown in Fig.~\hyperref[fig:A-T_warped_star]{\ref*{fig:fit_for_069}(a)}, the Y-axis represents the relative brightness of a burst compared to its preceding one, while the X-axis gives the recurrence interval between the previous burst and the current one, i.e., $t_{\rm n} - t_{\rm n{-}1}$.
In this diagram, each long/short-strong/weak pair occupies two opposite quadrants, and as the EMRI undergoes apsidal precession, successive pairs shift, thereby tracing a closed elliptical track.
The tilt (or slope) of the ellipses reflects how the warped disk's geometry\textemdash induced by LT torques from the SMBH spin\textemdash correlates burst brightness (energy) with recurrence time.

We now test the warped-disk predictions against observations, specifically the TDE-associated QPE source GSN~069. The observed data plotted in Fig.~\hyperref[fig:A-T_warped]{\ref*{fig:fit_for_069}(a)} display a clear negative correlation, which is well reproduced by the theoretical model in the stellar‑mass black hole scenario with $a<0$ (blue squares), corresponding to a negatively tilted elliptical trajectory in the BRD. In contrast, the $a>0$ scenario (red dots) predicts a positively tilted elliptical trajectory, while the flat‑disk ($a=0$) scenario (yellow triangles) yields a horizontally aligned ellipse (i.e., no clear correlation). 
The red and blue points are not perfectly symmetric. This asymmetry arises because the disk structure also depends on the direction of the accretion flow (which, in our setup, always maintains a small inclination relative to the EMRI orbit). Reversing the SMBH spin alone does not produce a perfectly mirrored disk configuration.

Likewise, a star secondary on a prograde orbit ($a>0$) could also reproduce the negative-correlation trend in the BRD (Fig.~\hyperref[fig:A-T_warped_star]{\ref*{fig:fit_for_069}(b)}), as the collision energy now scales positively with $v_{\rm rel}$ (Eq.~\ref{eq:collision_energy}). These two scenarios (a prograde star or a retrograde sBH) are nevertheless distinguishable in principle: in the BRD, the elliptical trajectory of the data points would be traced clockwise for the sBH case but counterclockwise for the star case. 
Future monitoring observations may potentially differentiate between these two cases. This is another potential application of the BRD.
These EM constraints on the prograde or retrograde nature and secondary composition of EMRIs set the stage for independent GW verifications (see below).

We also note three other QPEs with long-term monitoring: eRO‑QPE1, eRO‑QPE2, and RX~J1301.9+2747 \citep{2024ApJ...965...12C,2024A&A...690A..80A,2024A&A...692A..15G}. A preliminary overview of their BRDs reveals distinct behaviors: eRO‑QPE1 shows no clear correlation, suggesting a flat disk; eRO‑QPE2 exhibits a correlation similar to GSN~069, implying either a prograde star or a retrograde sBH; RX~J1301.9+2747, in contrast, displays an opposite correlation, pointing to a retrograde star or a prograde sBH. A detailed BRD analysis of these sources will be pursued in future studies.

\section{Long-term Light Curve}
\label{sec:ltlc}
Having matched the observed BRD, we next test whether our model can directly reproduce the long-term QPE light curve, taking the retrograde sBH scenario as an example. Figure~\hyperref[fig:Long-term]{\ref*{fig:fit_for_069}(c)} shows the fitted light curves for epochs XMM3, XMM4, Chandra, and XMM5 \citep{2023A&A...670A..93M}. We integrate the EMRI trajectory using post-Newtonian approximations up to 3.5PN order \citep{2023A&A...675A.100F,2014LRR....17....2B,2013CQGra..30g5017B}. The semi-major axis \(A\) and eccentricity \(e\) are used to set the initial conditions at pericenter, \(r_{\rm p}=A(1-e)\) and \(v_{\rm p}=\sqrt{(1+e)/r_{\rm p}}\) (in units \(G=M_\bullet=c=1\)). Each burst is approximated as a Gaussian peak, with luminosity \(L_{\rm peak}=E_{\rm QPE}/\sqrt{2\pi}\sigma\). For GSN~069, we adopt \(\sigma =800\,\mathrm{s}\).

The model accounts for the brightness--recurrence correlation and roughly reproduces the long-term light curve observed by XMM3--5 and Chandra, but it does not reproduce the observed decline in the \emph{absolute} flare amplitude. The fading may arise from secular effects not captured by the model, for example, damage to the star or draining of the disk due to repeated collisions \citep{2025arXiv250421456M}. Later QPEs (XMM6, XMM12, etc. \citep{2023A&A...674L...1M,2025A&A...693A.179M}) occurred during and after a significant X-ray re-brightening, likely associated with a second TDE, and are therefore excluded, our model nevertheless predicts that the brightness--recurrence correlation should persist.

The above fitting is intended as a simplified validation of our model. We acknowledge several simplifying assumptions: the EMRI orbit is taken to lie in the SMBH equatorial plane (a misaligned orbit would undergo Lense-Thirring nodal precession with a minimum period \(T_{\rm nod}\gtrsim2\,{\rm yr}\) for SMBH spin $|a| < 1.0$; see Eq.~\ref{eq:LT_frequency}); precession and alignment of the warped disk are omitted (see Sec.~\ref{sec:disk-derivation}); and we ignore relativistic light-travel delays as well as orbital perturbations from individual collisions. 

\label{sec:obs-data}
\label{SM:Obs}
The BRD uses XMM--Newton observations 3--5 (XMM3--5) and three QPE peaks from a single Chandra observation \citep{2023A&A...670A..93M}.
For XMM (Table~\ref{tab:sm-xmm}), each peak spectrum is fitted in the 0.2--1~keV band to give the bolometric peak luminosity $L_{\rm bol}$ and a Gaussian width $\sigma$; the collision-energy proxy is $E=\sqrt{2\pi}\,L_{\rm bol}\,\sigma$.
In these data, more luminous flares generally also last longer, so brightness and energy trends paint a similar picture.
Although calibration below 0.3~keV is imperfect, within-observation QPE ratios are largely unaffected.
Recurrence intervals $\Delta t$ are the differences between consecutive Gaussian peak times from fits.
For Chandra (Table~\ref{tab:sm-chandra}), peak spectra are photon-starved ($\lesssim$ few tens of photons), so we use $N_{\rm gauss}\,\sigma_{\rm gauss}$ from the light-curve fit as a proxy for $L_{\rm bol}\,\sigma$; these points are highly uncertain and are shown separately in the BRD.
All uncertainties are $1\sigma$.

\begin{table}[tbp]
\caption{\label{tab:sm-xmm}XMM3--5 QPE peaks for GSN~069.
$L_{\rm bol}$ (in $10^{42}\,\mathrm{erg\,s^{-1}}$) and $\sigma$ are from 0.2--1~keV peak spectral fits; $\Delta t$ (in seconds) is the recurrence interval from the preceding peak in the 0.4--1~keV light curve.}
\begin{ruledtabular}
\begin{tabular}{lccc}
Obs. & $L_{\rm bol}$ & $\sigma$ & $\Delta t$ \\
\hline
XMM3 & $3.07\pm0.14$ & $856\pm23$ & --- \\
     & $3.60\pm0.12$ & $886\pm17$ & $29757\pm40$ \\
\hline
XMM4 & $3.60\pm0.14$ & $922\pm27$ & --- \\
     & $2.89\pm0.13$ & $876\pm22$ & $33088\pm44$ \\
     & $3.18\pm0.13$ & $861\pm18$ & $31266\pm41$ \\
     & $2.87\pm0.13$ & $876\pm22$ & $32592\pm39$ \\
     & $3.52\pm0.14$ & $929\pm20$ & $31656\pm37$ \\
\hline
XMM5 & $2.65\pm0.17$ & $869\pm31$ & --- \\
     & $3.21\pm0.15$ & $981\pm22$ & $30081\pm46$ \\
     & $2.75\pm0.14$ & $870\pm23$ & $33459\pm41$ \\
     & $3.09\pm0.15$ & $971\pm23$ & $31181\pm42$ \\
     & $2.87\pm0.15$ & $993\pm28$ & $33373\pm46$ \\
\end{tabular}
\end{ruledtabular}
\end{table}

\begin{table}[tbp]
\caption{\label{tab:sm-chandra}Chandra QPE peaks for GSN~069 from a single observation.
$N_{\rm gauss}$ (counts~s$^{-1}$) and $\sigma_{\rm gauss}$ (seconds) are from the light-curve Gaussian fit; $\Delta t$ (seconds) is the recurrence interval from the preceding peak.}
\begin{ruledtabular}
\begin{tabular}{ccc}
$N_{\rm gauss}$ & $\sigma_{\rm gauss}$ & $\Delta t$ \\
\hline
$0.059\pm0.008$ & $720\pm55$ & --- \\
$0.034\pm0.006$ & $745\pm97$ & $33551\pm87$ \\
$0.040\pm0.006$ & $713\pm77$ & $31865\pm92$ \\
\end{tabular}
\end{ruledtabular}
\end{table}

\section{Gravitational-wave follow-up}
\label{sec:gw}
If QPEs originate from a stellar-mass black hole colliding with the disk, the embedded EMRI would become a unique viable target for LISA.
Although the EMRI in GSN~069 is expected to lie below LISA's detection threshold in the 2030s \citep{2025ARA&A..63..379K,2026ApJ...997..134Z,2026PASJ...78..185S},
EMRIs in several QPE sources exhibit orbital decay at rates exceeding the purely GW-driven contribution, probably driven by disk interactions \citep{2023A&A...670A..93M,2025ApJ...985..242Z,2024A&A...690A..80A, 2025arXiv250807961L}.
We can therefore anticipate a scenario in which an EMRI efficiently decays into LISA's detection band within $\sim 10~\rm yr$ under favorable disk–EMRI interaction-driven decay.

For instance, eRO-QPE2 is a promising GW follow-up candidate based on its electromagnetic properties \citep{2024A&A...690A..80A,2025ARA&A..63..379K,2026ApJ...997..134Z}. Its GW signal is expected to enter LISA's detection band in the 2030s, with a predicted signal-to-noise ratio ${\rm SNR} \sim 8.5–16.6$ for a standard stellar-mass secondary, and ${\rm SNR} \sim 28.8–69.7$ for a more massive one \citep{2026ApJ...997..134Z}. 

\begin{figure}[tbp]
\centering
\includegraphics[width=\columnwidth]{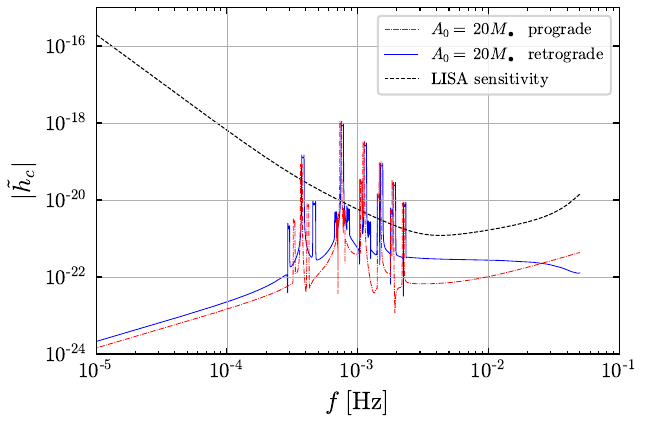}
\caption{
The characteristic strain over LISA’s four-year operational time for an EMRI system with sufficiently small semi‑major axis ($A_0=20M_\bullet$). The red and blue curves represent signals from prograde and retrograde EMRI systems. Here we set $M_\bullet=10^{5.99}M_\odot$, $m_{\rm sBH}=20M_\odot$, $e = 0.01$, SMBH spin $|a|=0.9$, $x_0=\cos I_0 =1.0$, $d_{\rm L} = 78.6~\rm Mpc$.
}
\label{fig:GW}
\end{figure}

As presented in Fig.~\ref{fig:GW}, we compute the characteristic strain of an EMRI with a $20\,M_\odot$ stellar-mass black hole at $A\sim20M_\bullet$ and $e\sim0.01$ over a four-year observing period using the \texttt{FastEMRIWaveforms} package \citep{Chua:2020stf,Katz:2021yft,Speri:2023jte}. The signal is detectable by LISA, and waveforms for $a=0.9$ and $a=-0.9$ are highly distinct, with a mismatch of \(\sim0.99\), indicating that a future gravitational-wave follow-up could cross-check the prograde/retrograde nature inferred from our BRD analysis, as well as other EMRI parameters. 
A successful cross-validation would represent a novel discovery: the first gravitational-wave follow-up originally identified through its electromagnetic signals. In such a system, electromagnetic signals would provide strong priors on key EMRI parameters, such as SMBH mass, spin, and sky localization, enabling accurate multimessenger parameter estimation and supporting precise tests across cosmology, astrophysics, and fundamental physics.

\section{Conclusion}
\label{sec:conclusion}
We have demonstrated that a warped accretion disk, formed in the late stage of a TDE, provides a natural framework for understanding the correlated long/short-strong/weak patterns observed in QPEs. The warp imprints the SMBH spin onto the recurrence and brightness features of the eruptions. Thus, the BRD reveals at least the prograde or retrograde nature of the EMRI orbit relative to the SMBH spin. Beyond this, BRD offers a potential means to distinguish between a star and a black hole secondary, based on its evolution direction along the elliptical track.

Our analysis of GSN~069 shows that the observed BRD pattern is consistent with either a prograde star or a retrograde stellar-mass black hole secondary. While current electromagnetic data alone cannot break this degeneracy, future high-cadence, long-term X-ray monitoring may. A preliminary inspection of other long-term monitored QPEs (eRO-QPE1, eRO-QPE2, and RX~J1301.9+2747) reveals distinct BRD behaviors, suggesting a diversity in their underlying EMRI–disk configurations.

If the secondary is a stellar-mass black hole, the embedded EMRI could become a target for LISA. 
We can anticipate a favorable case where disk–EMRI interaction-driven orbital decay brings the EMRI into the detectable band within a decade.
A future gravitational-wave follow-up would then provide a powerful test of our model, yielding precise multimessenger measurements of the EMRI system. This framework would position QPEs as electromagnetic precursors to LISA-detectable EMRIs, enabling statistical tests of SMBH spin distributions.

\begin{acknowledgments}
B.Y. is supported by Natural Science Foundation of China (NSFC) grants 12322307, 12361131579, and 12273026; by Xiaomi Foundation / Xiaomi Young Talents Program. G.M. acknowledges support from grant n. PID2023-147338NB-C21 funded by Spanish MICIU/AEI/10.13039/501100011033 and ERDF/EU. N.J. is supported by NSFC grants 12522303, 12192221. T.Y. is supported by the National Natural Science Foundation of China Grant No. 12575063. 
K.L. was supported by National Key R\&D Program of China (No. 2024YFC2207400).
X.-H. D. is supported by National Natural Science Foundation of China (Grant No. 12573017).
Z.-H. Z. is supported by the National Natural Science Foundation of China under Grants Nos. 12433001 and 12021003. The data analysis in this paper have been done on the supercomputing system in the Supercomputing Center of Wuhan University. 
We thank Yi-Ming Hu, Bin Liu, Ya-Ping Li, Dong Lai, and Yan-Rong Li for helpful and inspiring discussions.
\end{acknowledgments}

\appendix

\section{Parameter survey: disk, tearing, and BRD}
\label{sec:sm-params}
\label{SM:Params}

\subsection{Parameter survey}
\label{sec:sm-fiducial}

To address the sensitivity of the warped-disk QPE solution (and the related question of disc tearing; see below),
we perform a one-dimensional parameter survey, varying in turn
$\theta_{\rm out}\in\{10^\circ,15^\circ,20^\circ,25^\circ,30^\circ\}$,
$\alpha\in\{0.03,0.05,0.1,0.15\}$,
$H/R\in\{0.004,0.0055,0.008,0.01,0.02\}$,
$r_{\rm out}/R_g\in\{500,750,1000,1500,2000\}$,
and the surface-density slope $n$ in $\Sigma\propto r^{n}$ with $n\in\{-0.5,-2/3,-0.8\}$.
Each case re-solves the steady warp eigenmode, evaluates the tearing diagnostic $\psi(r)$,
and recomputes a short sBH--disk light curve and BRD with the same EMRI integration settings as in the main analysis.

Disk masses inferred from the adopted $\Sigma(r)$ span $\sim0.6$--$3.7\,M_\odot$ across the grid
(fiducial $M_{\rm disk}\simeq1.5\,M_\odot$), assuming $\Sigma=10^5\,\mathrm{g\,cm^{-2}}$ at the collision site.
These $M_{\rm disk}$ values are compatible with the order-of-magnitude TDE stellar mass inferred for GSN~069 \citep{2018ApJ...857L..16S,2021ApJ...920L..25S,2023A&A...670A..93M}.

The inner aligned region is defined by $\theta<H/R$. Across the grid its outer edge lies at
$r_{\rm align}\simeq43$--$125\,R_g$ (fiducial $\simeq101\,R_g$), always inward of the EMRI semi-major axis $A=160.1\,R_g$,
so collisions occur outside the fully aligned zone and the warp still modulates the impact geometry.

The fiducial parameters are
$M_\bullet=10^{5.99}\,M_\odot$,
$a=-0.9$,
$r_{\rm out}=1000\,R_g$,
$\theta_{\rm out}=20^\circ$,
$H/R=0.0055$,
$\alpha=0.1$,
$\Sigma\propto r^{-2/3}$ with normalization $\Sigma=10^5\,\mathrm{g\,cm^{-2}}$ at the collision site, 
semi-major axis $A=160.1\,R_g$,
eccentricity $e=0.015$,
and secondary mass $m=10\,M_\odot$.

\newpage

\subsection{Disc-tearing diagnostic}
\label{sec:sm-tearing}

A strongly warped disc may become unstable and break into discrete annuli (``disc tearing'')
when the dimensionless warp amplitude

\begin{equation}
  \psi(r)\;\equiv\; r\left|\frac{\partial\mathbf{l}}{\partial r}\right|
\end{equation}

exceeds a critical value $\psi_c$ that depends primarily on the Shakura--Sunyaev viscosity $\alpha$
\cite{2018MNRAS.475..758D,2021ApJ...909...81R,2021ApJ...909...82R}.
Writing $\mathbf{l}=(\sin\theta\cos\phi,\sin\theta\sin\phi,\cos\theta)$ and unwrapping $\phi$ before differentiation gives the practical form

\begin{equation}
  \psi^{2}(r)
  \;=\;
  \bigl(r\,\partial_r\theta\bigr)^{2}
  \;+\;
  \bigl(r\sin\theta\,\partial_r\phi\bigr)^{2}.
\end{equation}

We evaluate $\psi(r)$ from the eigenmode $\theta(r)$ and $\phi(r)$.
There is no simple closed expression $\psi_c(\alpha)$.
We adopt $\psi_c\approx0.4$, $0.5$, $2.7$, and $3.6$ at $\alpha=0.03$, $0.05$, $0.1$, and $0.15$
(Fig.~5 in \citet{2018MNRAS.475..758D}).

\begin{figure}
\includegraphics[width=\linewidth]{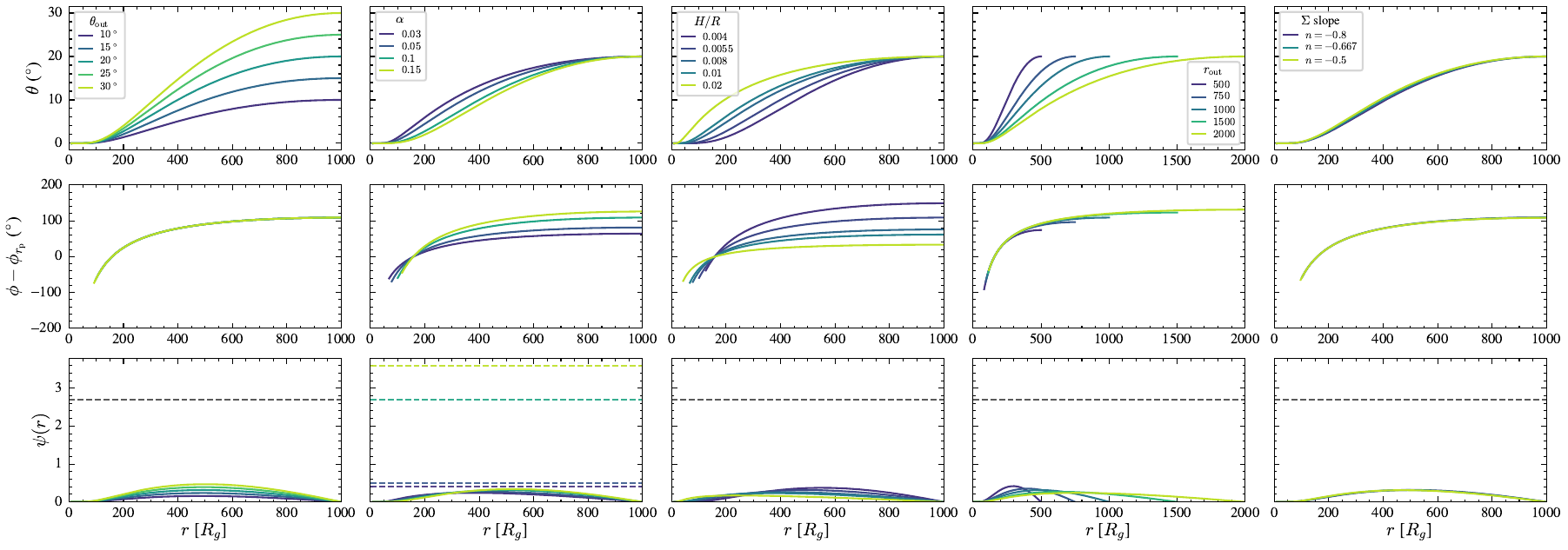}
\caption{\footnotesize Warped-disk structure under one-parameter variations of
  $\theta_{\rm out}$, $\alpha$, $H/R$, $r_{\rm out}$, and the surface-density slope $n$
  ($\Sigma\propto r^{n}$; $a=-0.9$).
  Within each column, solid curves of different colors show all surveyed values.
  Rows: inclination $\theta(r)$, twist $\phi(r)-\phi(r_{\rm p})$, and dimensionless warp amplitude $\psi(r)$.
  Dashed lines indicate $\psi_c$ from the local tearing analysis
  \cite{2018MNRAS.475..758D,2021ApJ...909...81R,2021ApJ...909...82R};
  in the $\alpha$ column, each $\psi_c(\alpha)$ uses the same color as the corresponding solid curve.
  All solutions remain below $\psi_c$.}
\label{fig:sm-disk-profiles}
\end{figure}

Figure~\ref{fig:sm-disk-profiles}
shows $\theta(r)$, $\phi(r)-\phi(r_{\rm p})$, and $\psi(r)$ for all five one-parameter sequences
($a=-0.9$; dashed: $\psi_c$, color-matched in the $\alpha$ column where $\psi_c=\psi_c(\alpha)$).
Across the survey, $\psi_{\rm max}<\psi_c$ in every case
(fiducial: $\psi_{\rm max}\approx0.31$ at $\alpha=0.1$, $\psi_c\approx2.7$).
Increasing $\theta_{\rm out}$ or decreasing $H/R$ or $r_{\rm out}$ sharpens the warp and raises $\psi_{\rm max}$,
but never reaches $\psi_c$; increasing $\alpha$ mainly enlarges $\psi_c$; while the surface-density slope $n$ has only a weak effect.

\subsection{Light curves and brightness--recurrence diagrams}
\label{sec:sm-lc-brd}

\begin{figure}
\includegraphics[width=\linewidth]{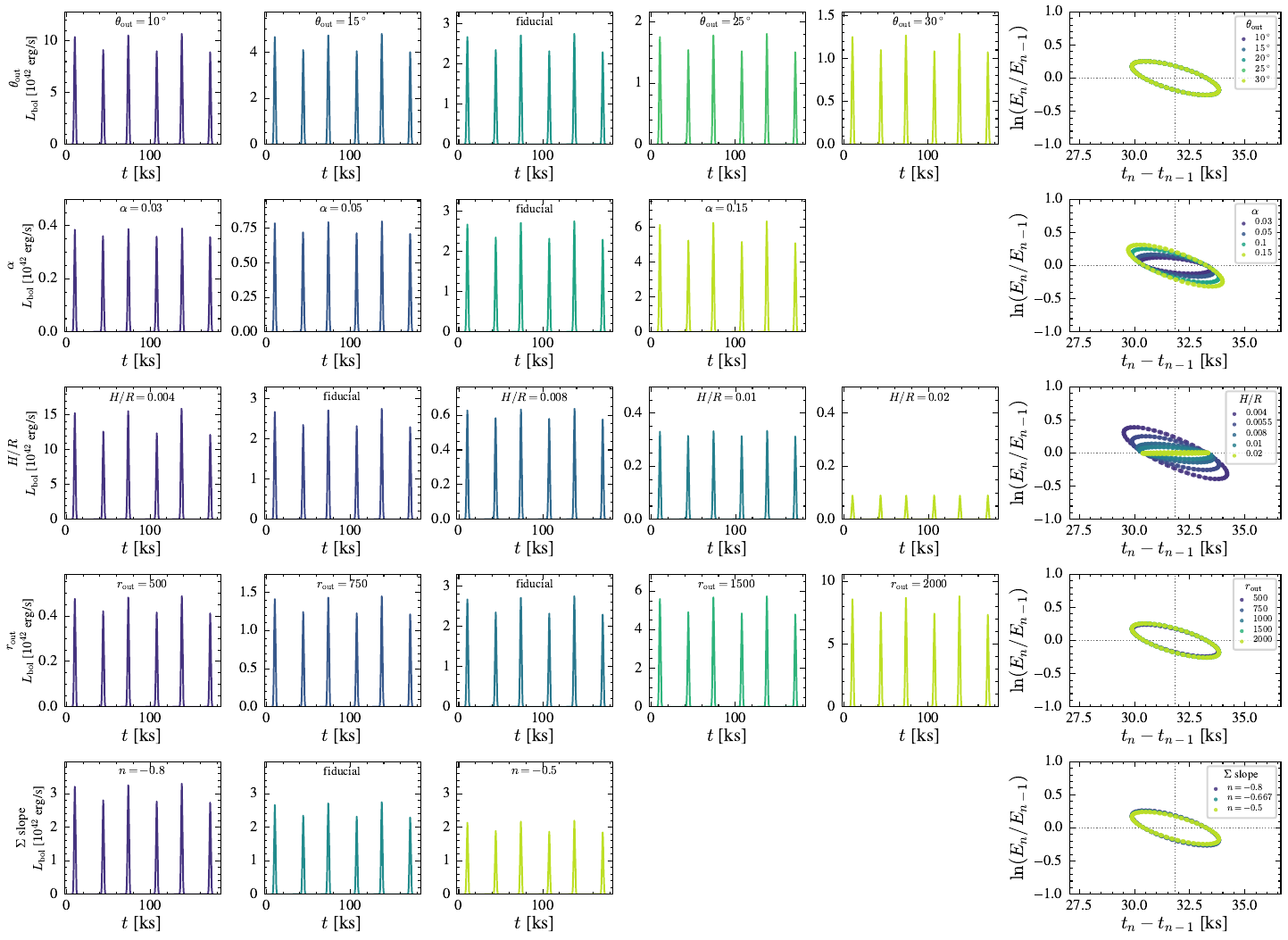}
\caption{\footnotesize Early sBH--disk light curves and overlaid BRDs for the five one-parameter scans
(rows: $\theta_{\rm out}$, $\alpha$, $H/R$, $r_{\rm out}$, $\Sigma$ slope $n$; fiducial included in each row).
Left columns: first six flares for each scanned value (independent $y$-scales when peak luminosities differ by more than a factor of $\sim2$).
Right column: BRD with all values of that axis overplotted in matching colors.
The tilted brightness--recurrence correlation persists across the full survey.}
\label{fig:sm-lc-brd-all}
\end{figure}

Figure~\ref{fig:sm-lc-brd-all} shows, for each scanned axis,
the earliest six flares (left panels) together with the corresponding BRD
(right panel; $\sim100$ successive bursts).
Varying $\theta_{\rm out}$ or $r_{\rm out}$ mainly changes the absolute peak luminosity,
with only a limited effect on the BRD morphology.
By contrast, $\alpha$ and $H/R$ reshape the BRD contour more noticeably;
the overall trend nevertheless remains a negative brightness--recurrence correlation,
except for the $H/R=0.02$ case,
so the inference of the SMBH spin orientation is unaffected.
The surface-density slope $n$ has only a mild impact on both the light curves and the BRD.

\clearpage
\onecolumngrid
\fixedfigures

\vspace{0.35em}

\clearpage
\twocolumngrid

\section{Energy-release scaling and the sense of BRD rotation}
\label{sec:energy-scaling}
\label{SM:Energy}

We relate the radiated burst energy to a power of the relative collision speed,
$E_\star\propto v_{\rm rel}^{2}$ for a stellar secondary and $E_{\rm sBH}\propto v_{\rm rel}^{-2}$ for a stellar-mass black hole.
Because the mapping from injected kinetic energy to the observed X-ray signal can be more complex,
we test how the BRD depends on this choice.
We adopt a one-parameter proxy $Q\propto v_{\rm rel}^{\beta}$ evaluated at each EMRI--disk collision
and plot $\ln(Q_n/Q_{n-1})$ against the recurrence interval $t_n-t_{n-1}$.
The disk and orbit are the fiducial warped-disk solution described above ($e=0.015$);
each trajectory is cut to a measured half apsidal cycle for $a=\pm0.5$ and $\pm0.9$.

\begin{figure}
\begin{minipage}[t]{0.48\textwidth}
\centering
\includegraphics[width=\linewidth]{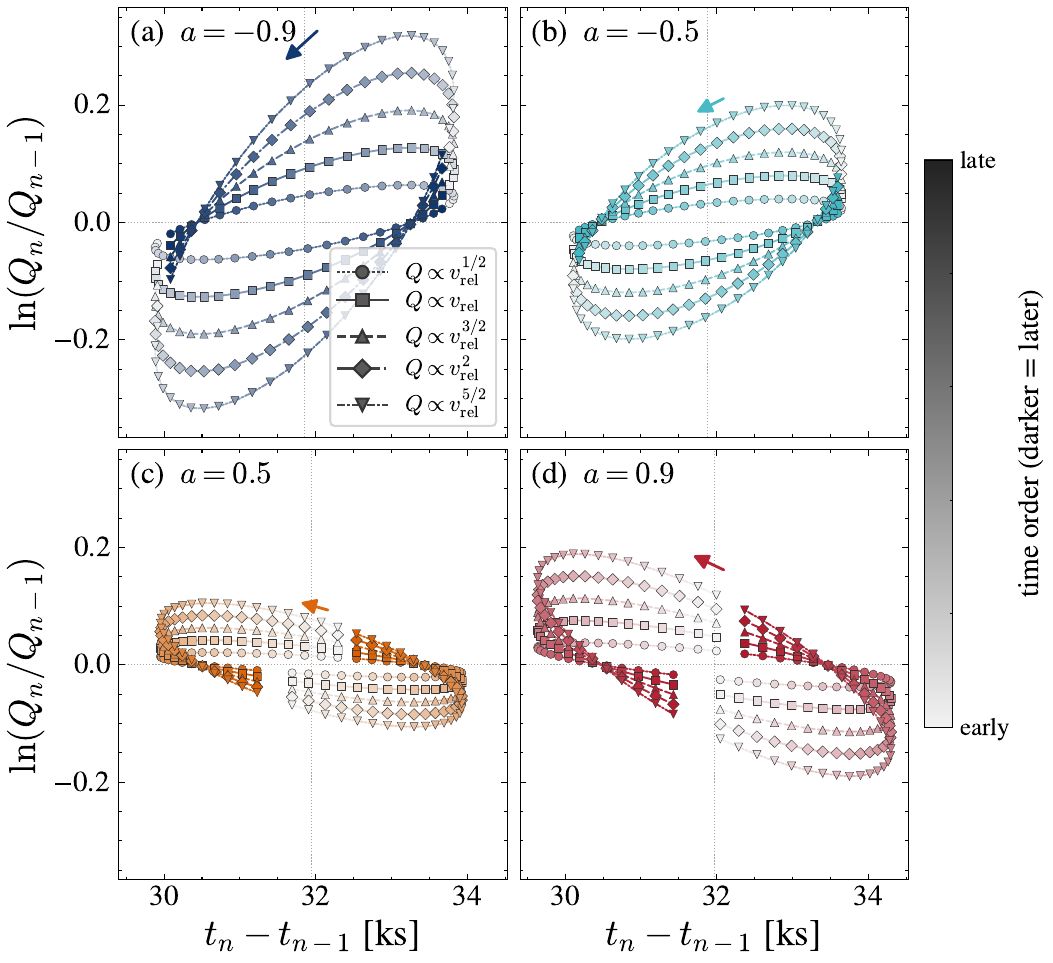}
\phantomsection\label{fig:sm-brd-beta}%
\text{(a) $Q\propto v_{\rm rel}^{\beta}$, $\beta>0$}
\end{minipage}\hfill
\begin{minipage}[t]{0.48\textwidth}
\centering
\includegraphics[width=\linewidth]{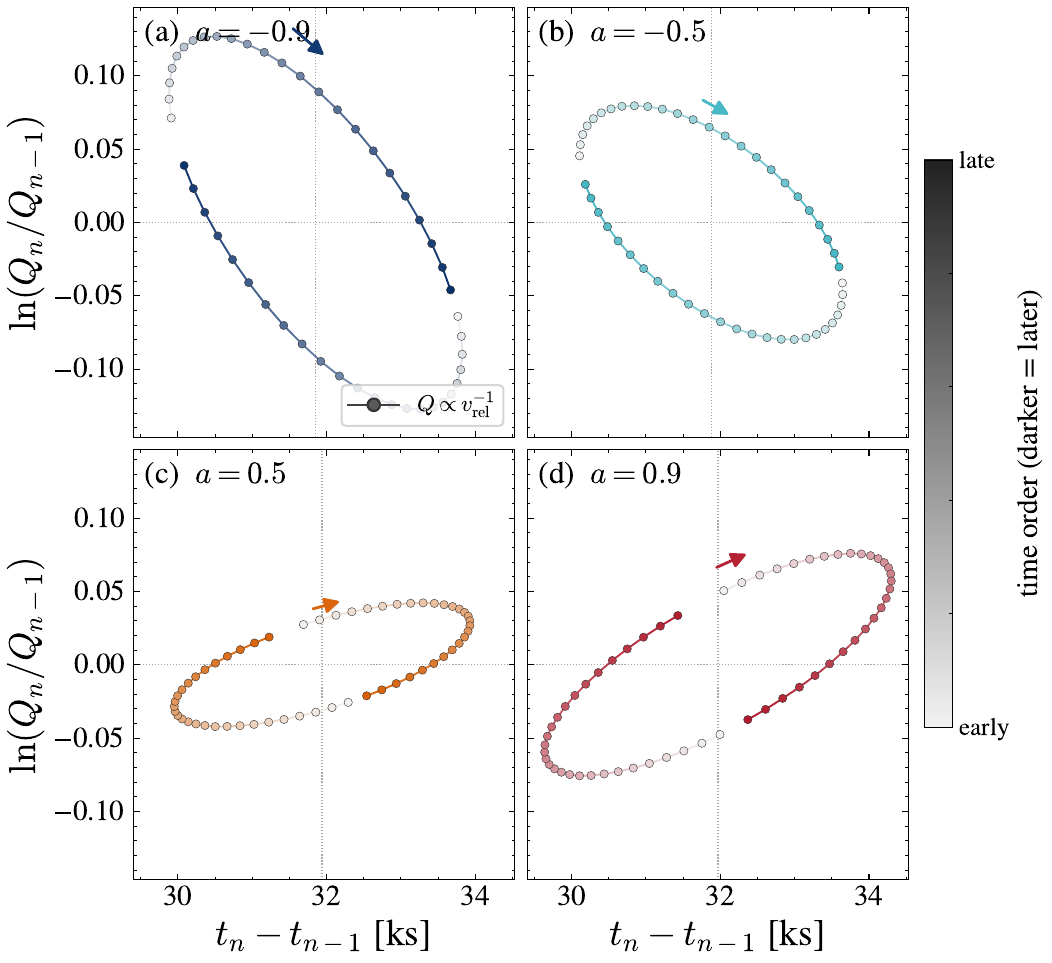}
\phantomsection\label{fig:sm-brd-betam1}%
\text{(b) $Q\propto v_{\rm rel}^{-1}$}
\end{minipage}
\caption{BRDs over a half apsidal cycle for $a=-0.9$, $-0.5$, $+0.5$, and $+0.9$
(panels a--d in each plot).
(a) Brightness proxy $Q\propto v_{\rm rel}^{\beta}$ with $\beta\in\{1/2,1,3/2,2,5/2\}$.
Colour encodes time order (darker = later).
Markers and linestyles distinguish the index $\beta$ (legend in panel a).
A positive $\beta$ only changes the vertical amplitude.
The tracks run counterclockwise; reversing $a$ reverses the tilt but not the sense of rotation.
(b) Same as (a), but for $Q\propto v_{\rm rel}^{-1}$.
Relative to the $\beta>0$ sequences, both the tilt and the sense of rotation reverse
(the tracks now run clockwise).}
\label{fig:sm-brd}
\end{figure}

Figure~\hyperref[fig:sm-brd-beta]{\ref*{fig:sm-brd}(a)} shows $\beta\in\{1/2,1,3/2,2,5/2\}$.
For $\beta>0$, changing the index only stretches the BRD vertically.
The tracks run counterclockwise for every spin:
reversing $a$ reverses the tilt of the ellipse
(negative $a$: a positive brightness--recurrence trend; positive $a$: the opposite)
but does not change the sense of rotation.
This is seemingly because the apsidal precession of the EMRI is always prograde (in the same sense as the orbit), independent of the sign of $a$.

If instead $\beta<0$ (Fig.~\hyperref[fig:sm-brd-betam1]{\ref*{fig:sm-brd}(b)}),
both the tilt and the sense of rotation reverse,
as required by $\ln(Q_n/Q_{n-1})=\beta\ln(v_n/v_{n-1})$:
the tracks run clockwise, and the brightness--recurrence trend at each $a$ flips as well.

Thus a sign change of $\beta$ reverses both the sense of rotation and the tilt,
whereas a sign change of $a$ reverses only the tilt.
If the BRD is measured well enough to read the sense of rotation,
that sense first determines $\mathrm{sign}(\beta)$;
the remaining overall trend then determines $\mathrm{sign}(a)$,
without assuming a specific radiative index.

\onecolumngrid
\vspace{0.6em}

\clearpage
\twocolumngrid

\section{Alternative collision models for strong/weak bursts}
\label{sec:sm-alt}
\label{SM:Alternatives}

The main analysis explains the brightness--recurrence correlation of QPE bursts as arising from EMRI collisions with a warped disk.
However, within the collision framework, other mechanisms have also been proposed for the alternating strong/weak pattern
\cite{2023ApJ...957...34L,2025ApJ...993..186H,2025ApJ...991..147L,2026arXiv260202656J}.
A natural question is whether these alternatives can also produce a sustained brightness--recurrence correlation.
We therefore examine three cases on a planar disk:

\begin{enumerate}
\renewcommand{\labelenumi}{(\arabic{enumi})}
\item
orbital eccentricity moves successive collisions to different radii;
\item
orbital eccentricity changes the relative collision speed;
\item
forward and backward collision plumes may be asymmetric, so flares look unequally bright from opposite sides of the disk.
\end{enumerate}

To test these three cases, we construct quasi-brightness--recurrence diagrams
pairing the recurrence interval $\Delta t$ with the corresponding physical quantities:

\newpage

\begin{enumerate}
\renewcommand{\labelenumi}{(\arabic{enumi})}
\item
the collision radius $R$;
\item
the relative speed $v_{\rm rel}=\lvert\mathbf{v}_{\rm EMRI}-\mathbf{v}_{\rm disk}\rvert$;
\item
a same/opposite-side diagnostic
$s=\mathrm{sign}(\mathbf{v}_{\rm EMRI}\cdot\hat{\ell})\,\mathrm{sign}(\hat{n}_{\rm obs}\cdot\hat{\ell})$,
where $s=+1$ ($s=-1$) when the EMRI velocity and the line of sight lie on the same (opposite) side of the disk.
\end{enumerate}

\begin{figure}
\includegraphics[width=0.90\textwidth]{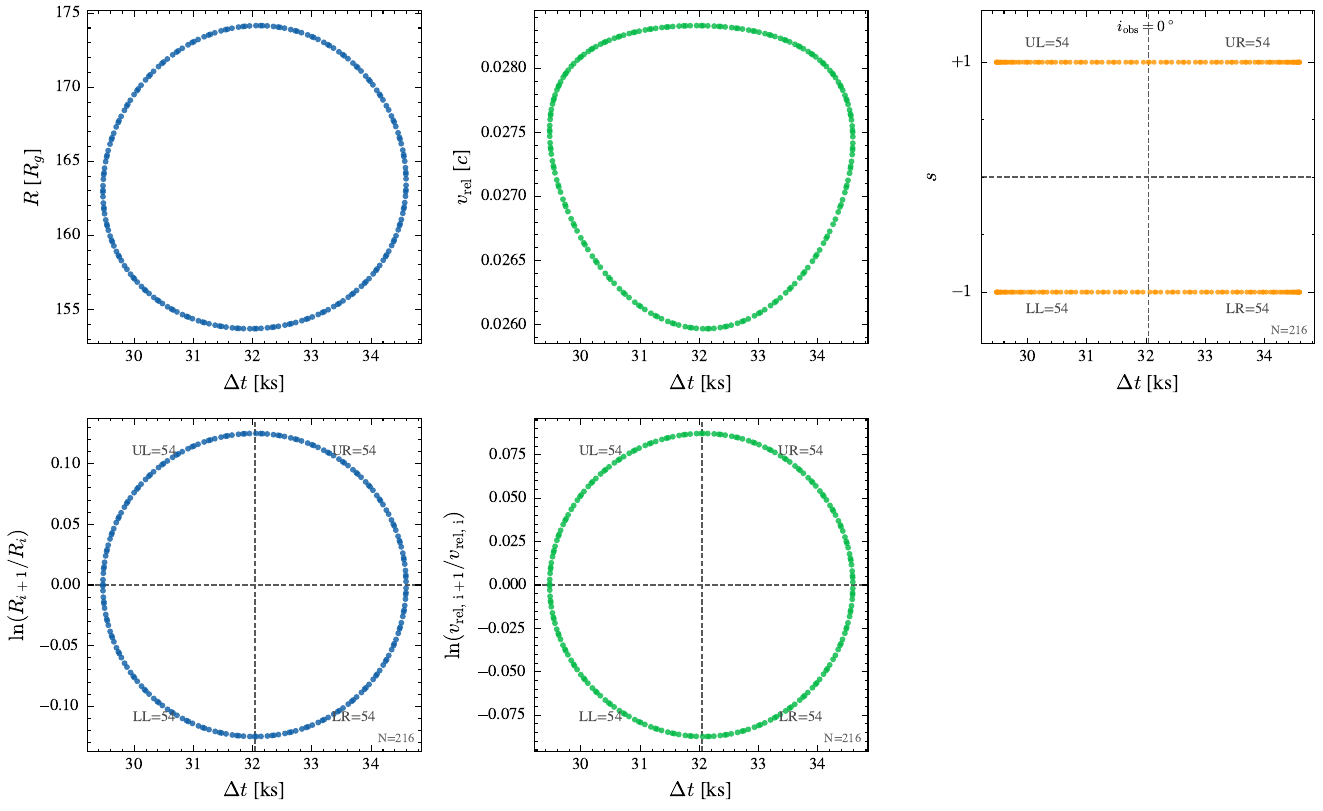}
\caption{quasi-BRDs for a flat disk ($e=0.04$).
The recurrence interval $\Delta t$ is plotted against
the collision radius $R$,
the relative impact speed $v_{\rm rel}$,
and the same/opposite-side indicator $s$.
Top panels show the absolute values of $R$, $v_{\rm rel}$, and $s$;
bottom panels show consecutive ratios.
Dashed lines and quadrant counts illustrate the absence of a preferred correlation.}
\label{fig:sm-flat-Rv}
\end{figure}

The calculations use a flat disk and $e=0.04$
\citep{2025ApJ...983L..18J},
and cover two apsidal precession periods.
Under a Newtonian potential, the fractional variation of the collision radius
(or likewise of $v_{\rm rel}$) between apoapsis and periapsis is of order
$(1+e)/(1-e)-1\sim 2e\sim10\%$ for $e=0.04$.
Figure~\ref{fig:sm-flat-Rv} shows the quasi-BRDs for $R$, $v_{\rm rel}$, and $s$.
In none of these cases do we find a clear correlation with the recurrence time.
The warped-disk model remains the only model that can produce a sustained brightness--recurrence correlation.

We also note that two observable flares per orbit need not apply to every QPE source
\citep{2026ApJ...1003..148A,2026arXiv260300226L,2025A&A...703A.263H,2025ApJ...978...91Y}.
A single flare per orbit would still occur in a different collision geometry if the disk is warped rather than flat, but it would not by itself produce the paired, alternating strong/weak and long/short pattern that motivates this work.
We therefore do not discuss one-flare-per-orbit sources further here.

\onecolumngrid
\vspace{0.6em}

\clearpage
\twocolumngrid
\restorefigures
\bibliographystyle{apsrev4-2}
\bibliography{apstemplate}

\end{document}